\documentclass[aps,prl,twocolumn,superscriptaddress]{revtex4}

\usepackage{graphicx,epsfig,amsmath,amssymb,amsfonts,mathrsfs,psfrag,color}

\begin{document}

\title{Shot Noise Signatures of Charge Fractionalization in the $\nu=2$ Quantum Hall edge}

\author{Mirco Milletar\`i }
\email[Current contact e-mail: ]{phymirc@nus.edu.sg}
\affiliation{Institut f\"ur Theoretische Physik, Universit\"at Leipzig, Br\"uderstr. 14, D-04103  Leipzig, Germany}
\affiliation{Max-Planck-Institute for Solid State Research, Heisenbergstr. 1, D-70569 Stuttgart, Germany}

\author{Bernd Rosenow}
\affiliation{Institut f\"ur Theoretische Physik, Universit\"at Leipzig, Br\"uderstr. 14, D-04103 Leipzig,  Germany}

\date{September 11, 2013}

\begin{abstract}

We  investigate the  effect of  interactions  on shot noise in $\nu=2$ quantum Hall edges, where 
a  repulsive coupling between co-propagating edge modes is expected to give rise to charge fractionalization.  Using the method of non-equilibrium bosonization, we find that even asymptotically the edge distribution function depends in a sensitive way on the interaction strength between the edge modes. We compute shot noise and  Fano factor from the asymptotic distribution function,  and from comparison with a reference model of fractionalized excitations we find that the Fano factor can be close to the value of the fractionalized charge. 

\end{abstract}


\maketitle

In contrast to three spatial dimensions, where excitations of an interacting many particle system often carry the same quantum numbers as in the non-interacting case, interactions in 1d systems completely change the character of the excitation spectrum \cite{Giamarchi,delft}. A prototype model for this physics is the Luttinger model, where electrons are no longer well defined quasi-particles, and where electronic excitations decompose into spin and charge parts moving with different velocities \cite{Giamarchi,Auslaender+05}.

An important example of interacting 1d systems are the edge states  of incompressible quantum Hall liquids 
\cite{Halperin82,Wen1990}, where as a result of strong interactions charge fractionalization can occur \cite{Safi+95,Berg+09,Neder12,Pham+00,Steinberg+08,Leinaas+09,Horsdal+11} and manifests itself in shot noise \cite{KaFi94,picciotto97,Siminadayar+97,Trauzettel04}.
For the case of filling fraction $\nu=2$, there are two chiral edge modes co-propagating at different velocities $v_1$ and $v_2$. In the presence of a short range interaction $v_{12}$ between them, a pulse of charge $e$ injected into edge mode one at a first quantum point contact (QPC1) decomposes into a charge pulse and a neutral pulse. In the charge pulse, a charge $e^* = (e/2) \sin 2\theta$ (where $\tan2\theta=v_{12}/(v_1-v_2)$ parametrizes the strength of interactions) travels on mode two and  $e/2 + \sqrt{e^2/4 -  (e^*)^2} $ on mode one \cite{supplemental}. In the neutral pulse, there is a charge $- e^*$ on mode two and a charge $e/2 - \sqrt{e^2/4 -  (e^*)^2} $ on mode one. In this way, by exciting edge channel one via a partially transmitting QPC1, high frequency charge noise is generated on edge mode two \cite{Berg+09}. At QPC2, allowing for partial transmission of channel two, both charges $\pm e^*$ traveling within the charge (neutral) pulse give rise to low frequency shot noise with a Fano factor $e^*/e$ \cite{supplemental} .

\begin{figure}[t]
\includegraphics[width=1.0\columnwidth]{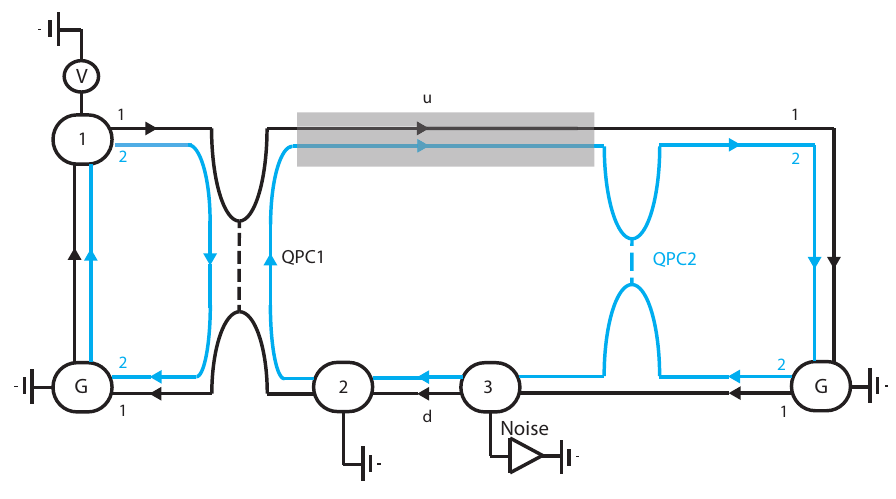}
\caption{(color online) Sketch of a $\nu = 2$ Hall bar with a QPC1, where inner  modes ($"2"$, light blue) are fully reflected, 
while partial transmission of outer modes ($"1"$,  black) is possible.  At  QPC2, the opposite situation is realized. The shaded area is the interaction region, where partial energy relaxation takes place. The upper edge is biased with voltage $V$ at contact 1, current noise is measured at contact 3.}
\label{setup}
\end{figure}

Alternatively, one can look at this problem by using the concept of energy relaxation \cite{Sueur+2010,Degiovanni+10,KoCh,LeSu12}. 
Interactions play a crucial role in the thermalization process that drives a system through states described by the Gibbs equilibrium ensemble. Generically,  the dynamics is only constrained by two integrals of motion,  total energy and total particle number. Integrable models like the $\nu=2$ quantum Hall edge
have infinitely many integrals of motion, and  therefore it is not clear if an equilibrium state can ever be reached \cite{RiDuOl}. 
If the two edge modes are driven out of equilibrium with respect to one another, the system relaxes towards a non-thermal steady state \cite{Sueur+2010,Degiovanni+10,KoCh,LeSu12, ScBaMi}, whose distribution function determines shot noise at  QPC2. 
The corresponding Fano factor  depends on the strength of the interaction between the edge modes, and in general neither agrees with the fractional charge $e^*$ introduced above, nor with the result for two equilibrated edge modes. For the special case of a half open QPC1 however, the Fano factor is close to $e^*/e$, suggesting an interpretation in terms of charge fractionalization. 
Some of our results were obtained independently in \cite{Neder12,LeSu12a}. In \cite{Neder12}, 
a setup similar to that in Fig. \ref{setup} was analyzed perturbatively in the transmission probability $a$ of QPC1, capturing only the initial stage of relaxation.  A non-perturbative analysis   is presented in \cite{LeSu12a}, and  the non-analytic dependence of noise on  $a$  in the limit  $a \ll1$ is emphasized. 
If integrability of the $\nu=2$ edge is broken, the system eventually relaxes to a thermal state \cite{Levchenko12}. 
%
%

We consider the setup  Fig.~\ref{setup} where a Hall bar is pinched by two QPCs. 
The outer edge mode is labeled  "$1$" and the inner one "$2$". The top and bottom edges originate  at zero temperature from  reservoirs at voltages $V_1=V$ and $V_2=0$.   
At QPC1, the outer modes are partially transmitted with probability $a$, while the inner ones are fully reflected; as a consequence, only the outer mode become noisy. After QPC1,  the two edge modes interact over some distance  (shaded area in Fig.~\ref{setup}) before reaching QPC2. Here, the outer modes are fully transmitted while the inner ones are partially reflected with probability $p$. Current noise is then measured at contact 3. 
Using the recently developed non-equilibrium bosonization technique \cite{Neder08,LeSu09,GuGeMi,PrGuMi} within a quantum-quench model \cite{Ca,IuCa,SoCaCh}, we compute the shot noise at QPC2, with particular emphasis on its dependence on the strength of the interaction between the edge modes. 

The edges and QPC2 are described by the following Hamiltonian ($\hbar=k_B=1$): 
%
\begin{eqnarray} \label{ham}
& & \mathscr{H}_{\rm \eta} = 2\pi \int_x \left(v_1 \rho_{1\eta}^2(x) +v_{2} \rho_{2\eta}^2(x) + v_{12} \rho_{1\eta}(x) \rho_{2\eta}(x) \right)  \nonumber  \\ 
& & \mathscr{H}_{\rm QPC2} =  t_2 \psi^{\dagger}_{2u}(x) \psi_{2d}(x) + h.c.  \label{qpc}
\end{eqnarray}
%
Here,  $\mathscr{H}_{\eta}$ describes  chiral modes, $\eta=u,d$ labels the upper and lower edge. The local interaction needs to satisfy the stability criterion $v_{12}^2/4 \leq v_1 v_2$ \cite{braggio}. $\mathscr{H}_{\rm QPC2}$ describes tunneling of electrons at QPC2  with $t_2$  the tunneling amplitude. The fields $\rho_{i\eta}(x)$ in (\ref{ham}) describe density fluctuations and are related to  bosonic displacement fields by $\rho_{i\eta}(x)=\partial_x\phi_{i\eta}(x)/2\pi$; here "i" labels different edge modes. The bosonic fields satisfy $\left[ \phi_{i\eta}(x),\phi_{j\xi}(y) \right] = \imath\, \pi\, \delta_{i\eta,j\xi} \, {\rm sign} (x-y)$, and the  fermionic field is represented as $\psi_{i\eta}(x)= ( 2\pi \alpha)^{-1/2} \,  e^{ \imath \phi_{i\eta}(x)}$ with $\alpha$ denoting a short distance cutoff on the scale of the magnetic length. For later reference, we decompose the bosonic fields as $\phi_{i\eta}(x)=\varphi_{i\eta}(x)+\varphi^{\dagger}_{i\eta}(x)$, $\varphi_{i\eta}(x)=\sum_{q>0} \sqrt{2\pi/q L} e^{-q \alpha/2} e^{\imath s_{\eta}q x}b_{i\eta}(s_{\eta} q), $ where $s_{\eta}=\pm 1$ respectively for right (u) and left (d) movers, and  $b^{\dagger}$, $b$  are canonical bosonic  operators.

Following \cite{KoCh}, we do not model QPC1 explicitly but instead consider its effect on the downstream electron distribution of mode $(1u)$ in a non-interacting setting, and model the distribution as a "double step" function
%
\begin{equation}\label{distr}
f(\epsilon) = a \, \theta(-\epsilon+\mu_1) + (1-a) \, \theta(-\epsilon+\mu_2),
\end{equation}
%
where $\mu_1=(1-a) e V$ and  $\mu_2=-a\, e V$ ($e V>0$) are chosen such that the average density in 
mode $(1u)$ corresponds to zero bias. As a consequence of this choice, there is no density shift in mode $(2u)$.

Next, we consider the effects of the inter-mode interaction on the  distribution function (\ref{distr}). Instead of switching on the interaction right after QPC1, 
we use the model of a quantum quench, where  the interaction $v_{12}$  is suddenly turned  on for times $t>0$ everywhere in space.
Due to the chirality of the edge states, the quantum quench faithfully models the effect of a position dependent interaction, see \cite{KoCh} and \cite{supplemental}.
The interacting Hamiltonian can  be diagonalized by means of a Bogoliubov transformation $M$. For co-propagating states ($v_1 v_2 >0$), $M$ can be represented by the following matrix:
%
\begin{equation} \label{bog}  
M=\left(\begin{array}{cc}\cos\theta & \sin\theta \\\ -\sin\theta & \cos\theta\end{array}\right), 
\end{equation}
%
%
\begin{figure}[t]
\includegraphics[width=0.85\columnwidth]{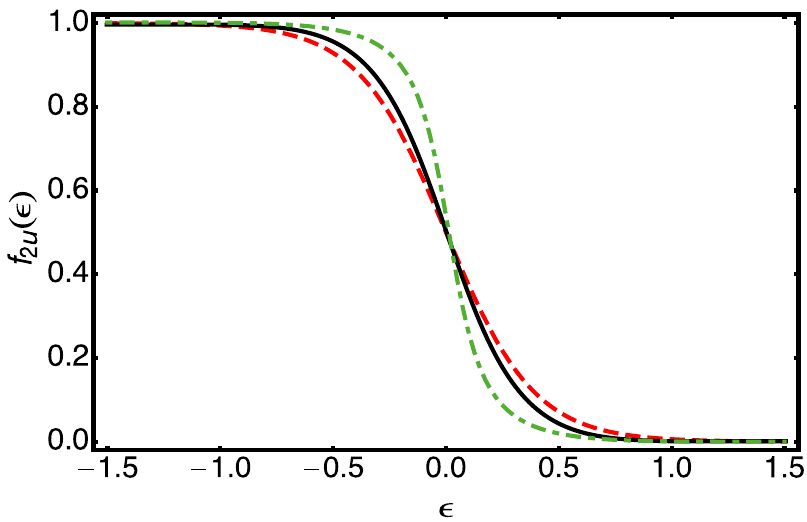}
\caption{ (color online) Steady state distribution of edge mode $(2u)$ asymptotically  away  from QPC1. (black full line) 
Non-equilibrium distribution obtained from  Eqs.~(\ref{bos},\ref{fredh}) by considering all cumulants. (green dash-dotted line) Distribution obtained by retaining only the gaussian term. (red dashed line) Fully equilibrated distribution at effective temperature $T^{*}=eV\sqrt{(3/2) a(1-a)}/\pi$. The mixing angle is $\theta=0.47$ and the transmission probability of QPC1 is $a=1/2$. }
\label{distrF}
\end{figure}
%
allowing to express $\mathscr{H}$  in terms of new fields $\beta_{i,q}=\sum_j M_{ij} b_{j,q}$. 
The mixing angle $\theta$ expresses the strength of the interaction through the relation $\tan2\theta=v_{12}/(v_1-v_2)$.
At this point the new operators evolve in the Heisenberg picture as $\beta_{iq}(t_0)=e^{-\imath q \tilde{v}_i t_0 } \beta_{iq}(t=0)$, with new velocities $\tilde{v}_{1(2)}  =  v_{1(2)} \cos^2\theta + v_{2(1)} \sin^2 \theta \pm \frac{1}{2} v_{12} \sin2\theta $. 

As a final step, we undo the Bogoliubov transformation in order to express the $\beta_{iq}(t_0)$ in terms of the original basis. As a result, we  obtain a relation between the bosonic operators at $t_0>0$ and those at $t=0$: 
%
\begin{eqnarray}\label{bosbog}
b_{1q}(t_0)=u_q(t_0) b_{1q}+s_q(t_0) b_{2q} \\ \nonumber
b_{2q}(t_0)=s_q(t_0) b_{1q}+v_q(t_0) b_{2q} 
\end{eqnarray}
%
where $b_{iq}\equiv b_{iq}(t=0)$. Now all the time dependence is encoded in the coefficients 
%
\begin{eqnarray} \label{coeff}
u_q(t_0)&=&\cos^2\theta \, e^{-\imath q \tilde{v}_1 t_0}+\sin^2\theta \, e^{-\imath q \tilde{v}_2 t_0} \\ \nonumber
v_q(t_0)&=&\cos^2\theta \, e^{-\imath q \tilde{v}_2 t_0}+\sin^2\theta \, e^{-\imath q \tilde{v}_1 t_0} \\ \nonumber
s_q(t_0)&=&\frac{1}{2} \gamma_\theta (e^{-\imath q \tilde{v}_1 t_0}-e^{-\imath q \tilde{v}_2 t_0} ), 
\end{eqnarray}
%
where $\gamma_{\theta}=\sin 2\theta$. To leading order in the tunneling amplitude $t_2$, the current noise at QPC2 can be expressed in terms of greater (lesser) Green functions $G_{i,\eta}^{> (<)}(\epsilon)$ \cite{Lev}  as
%
\begin{eqnarray}
S_{\omega \to 0} \! &=& \! {2 e^2\over h} {|t_2|^2 \over 2 \pi}   \! \int_{\epsilon} \! G_{2u}^{<}(\epsilon) G_{2d}^{>}(\epsilon) + G_{2d}^{<}(\epsilon) G_{2u }^{>}(\epsilon)  ,  
\label{noise}
\end{eqnarray}
%
with $G^{<}(\epsilon)=G^{>}(-\epsilon)$. Using the boson representation of 
electron operators,  we can compute   $G_{2u}^{> (<)}(\tau)$ of the fully interacting edge mode. Due to the non-equilibrium distribution of edge mode $(1u)$, calculating the expectation value of a product of bosonic exponents is highly non trivial. Here we discuss the results for the "long time limit" of the Green function, in which the system reaches a non-equilibrium steady state:   
%
%
\begin{figure}[t]
\includegraphics[width=0.85\columnwidth]{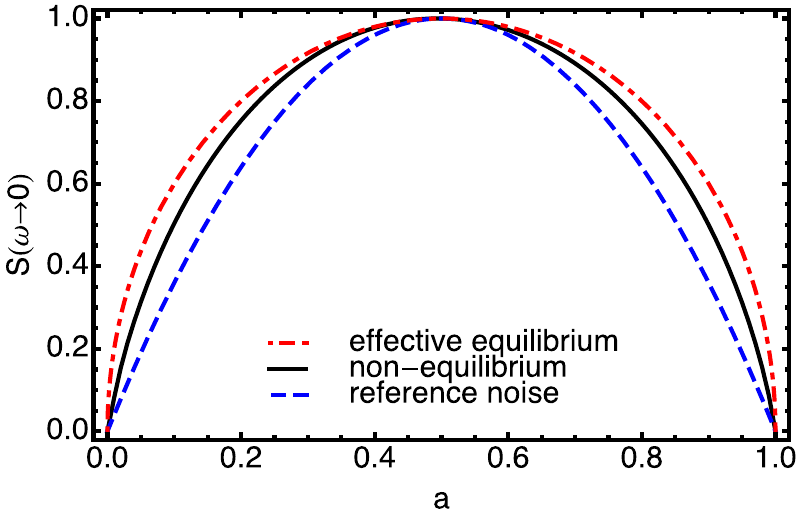}
\caption{(color online) Shot noise  after QPC2 as a function of $a$, normalized to its $a=1/2$ value, for a mixing angle $\theta=0.47$. (full black line) Full non-equilibrium result. (dashed blue line) Reference noise of non-interacting electrons. (dash-dotted red line) Noise in a fully equilibrated thermal state. }
\label{exnoise}
\end{figure}
%
\begin{eqnarray}\label{bos}
G_{2u}^{<}(\tau)&=& \langle \psi^{\dagger}_{2u}( t_0 +\tau,x_0) \psi_{2u}(t_0,x_0) \rangle \\ \nonumber
&=& G_{0}^{<}(\tau) \langle e^{\sum_{q} \lambda_{1u}^{\star}(q,t_0, \tau) b_{1u,q}^{\dagger}} e^{-\sum_{q} \lambda_{1u} (q,t_0, \tau) b_{1u,q} } \rangle. \\ \label{bos2}
G_{0}^{<}(\tau)&=& \frac{1}{2\pi} \frac{1}{ (-\imath \,\tilde{v}_1 \,\tau + \alpha)^{\sin^2\theta} } \frac{1}{(-\imath \, \tilde{v}_2 \, \tau + \alpha)^{\cos^2\theta}}. \nonumber
\end{eqnarray}
%
Here, $G_{0}^{<}(\tau)$ is the equilibrium Green function of edge mode 2 in the presence of interactions. All the information about non-equilibrium effects is contained in the average over bosonic coherent states in Eq.~(\ref{bos}), where $\lambda_{1u}(q,t_0,\tau)= \imath \, (2 \pi /q L )^{1/2} e^{\imath q x_0 - q \alpha/2 } [ s_q(t_0+\tau)-s_q(t_0)]$. As emphasized in \cite{GuGeMi}, non-equilibrium effects make the theory non-Gaussian,  and higher order cumulants appear in the evaluation of the above expectation value.  In order to compute the expectation value over the non-equilibrium state, we  refermionize the bosonic  operators introducing new fermionic  operators   \cite{delft}:
%
\begin{eqnarray}\label{ref}
b_{1u,q}^{\dagger} &=&  \imath (2 \pi /q L )^{1/2} \displaystyle \sum_{k} c_{1u,k+q}^{\dagger} c_{1u,k}  \ \ , \\ \nonumber
b_{1u,q} &=&  -\imath (2 \pi /q L)^{1/2} \displaystyle \sum_{k} c_{1u,k-q}^{\dagger} c_{1u,k} \ \ .
\end{eqnarray}
%
%
\begin{figure}[t]
\includegraphics[width=0.85\columnwidth]{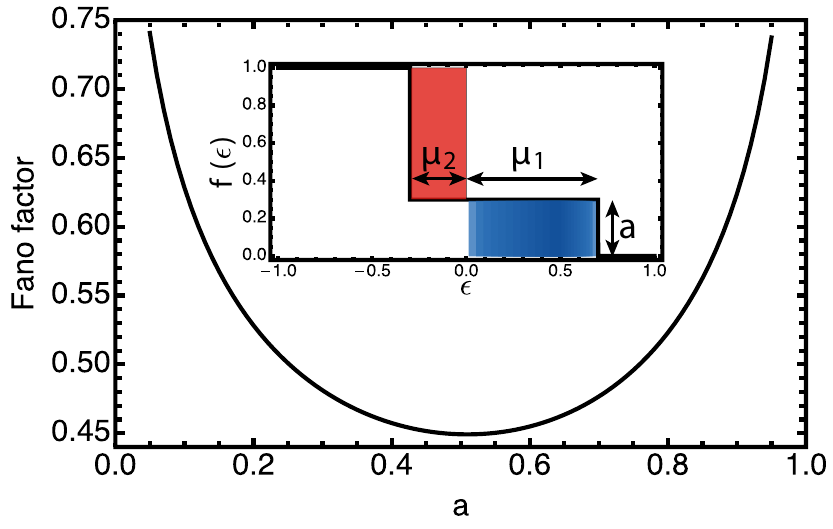}
\caption{(color online) Fano factor $F=S/S_{\rm ref}$ as a function of transparency of QPC1 for a mixing angle $\theta=0.47$. At $a=1/2$ the Fano factor is $F=0.45$. Inset : double step distribution Eq.~(2). Red area describes a hole current $I_h$  and blue area a particle current $I_p=I_h=(e^2/h) V a(1-a)$ impinging on QPC2. From this, we obtain a  reference noise  $S_{\rm ref}=2 \,e\, p \, (I_h+I_p)$, see also Eq.~(\ref{fano}). }
\label{fanopl}
\end{figure}
%
Since the bosonic operators describe free particle-hole excitations, also the $c$-operators are free and therefore can be connected to the incoming states via a scattering matrix. Then, the expectation values of products of 
Fermi operators can be evaluated using an appropriate fermionic density matrix $\rho_{1u}$. The crucial step now consists in noticing that the computation of higher order cumulants is similar to the problem of full counting statistics, and using Klich's trace formula \cite{PrGuMi,Kl} it can be expressed in terms of a Fredholm determinant of the Toeplitz type,  normalized to its zero temperature, equilibrium value 
%
\begin{equation} \label{fredh}
\bar{\Delta}_{\tau}(\delta)=\frac{\det \left[1+(e^{-\imath \delta_{\tau}}-1)f(\epsilon)\right]}{\det\left[1+(e^{-\imath \delta_{\tau}}-1)\theta(-\epsilon)\right]} \ \ , 
\end{equation} 
%
where $f(\epsilon)$ is given by Eq.~(\ref{distr}). The scattering phase $\delta_{\tau}=-\sum_q  (2 \pi /q L )^{1/2} [\lambda(q,t_0,\tau)+\lambda^*(q,t_0,\tau)]=2\pi (e^*/e) \omega_{\tau}(t_0,x_0)$ contains informations about the inter-edge interaction, and the window function 
%
\begin{eqnarray}\label{wind} \nonumber
\omega_{\tau}(t_0,x_0)&=&  \theta[x_0- \tilde{v}_1(t_0+ \tau) ]-\theta [x_0- \tilde{v}_1t_0 ] \\
& & + \theta [x_0- \tilde{v}_2 t_0]-\theta[x_0- \tilde{v}_2 (t_0+\tau)]  .
\end{eqnarray}
%
As a function of $t_0$, $\omega_{\tau}(t_0,x_0)$ represents two unit square pulses of opposite signs, with widths $\tau$, and with a separation equal to $x_0 (\tilde{v}_1^{-1}-\tilde{v}_2^{-1})$. Since $\delta_{\tau}=2\pi (e^*/e) \omega_{\tau}(t_0,x_0)$, these pulses can be identified with charges $\pm e^*$ passing an observer at position $x_0$. 
In the case of two separated pulses, the expectation value of bosonic coherent states  factorizes into a product of two single pulse determinants having the same scattering phase $\delta_{\tau,\rm single}=2\pi (e^*/e) [\theta(-t_0)-\theta(-t_0-\tau)]$, and we can rewrite Eq.~(\ref{bos})  as  $G_{2u}^{<}(\tau)=G_{0}^{<}(\tau) \, \bar{\Delta}_{\tau}^2(\delta_{\rm single})$. The determinant  Eq.~(\ref{fredh}) can be evaluated numerically by treating $t_0$ and $\epsilon$ as conjugated variables and by carefully defining a regularization scheme \cite{PrGuMi}. Finally, the lesser Green function $G_{2d}^{<}(\epsilon) =  \theta(-\epsilon)/\tilde{v}_1^{\sin^2\theta}\tilde{v}_2^{\cos^2\theta} $ is easily evaluated due to its equilibrium nature. Fourier transforming Eq.~(\ref{bos})  into energy space, we can compute the distribution function at QPC2;  
as a consequence of interactions, the distribution function is broadened from a single step (see Fig.~\ref{distrF}). However, it  does  not have the same functional form as a Fermi distribution, but rather describes a  non-equilibrium steady state. 
The distribution obtained by only retaining the Gaussian term in the cumulant expansion clearly deviates from the full one, 
making evident the necessity for including higher order terms. The non-equilibrium distribution also deviates from an equilibrium Fermi distribution with  effective temperature $T^{*}=eV\sqrt{(3/2) a(1-a)}/\pi$, obtained by assuming that the two edge modes fully equilibrate and that each of them carries half the energy flux injected into the upper edge via QPC1.  
Using Eq.~(\ref{noise}) we can finally evaluate the low frequency noise; in doing so we relate the reflection probability $p$ to the microscopic Hamiltonian trough $p= |t_2|^2/2 \pi \,\tilde{v}_1^{\sin^2\theta} \, \tilde{v}_2^{\cos^2\theta}$. In Fig.~(\ref{exnoise}), we display the dependence of  low frequency noise on the transmission $a$ of QPC1, normalizing the noise by its value at $a=1/2$. One clearly sees that it deviates both from the standard free fermion dependence $a(1-a)$, and from the effective equilibrium result with $S_{\rm eq}=4\, e\, p\, I\, \log2 \, \sqrt{(3/2)a(1-a)}/\pi $. To put the strength of the noise at QPC2 in perspective, we define a  reference noise expected for non-interacting electrons tunneling through both QPC1 and QPC2 along a single edge, obtained by using the distribution Eq.~(2) in Eq.~(6)
%
\begin{equation}\label{fano}
S_{\rm ref}(\omega \to 0)=4\, e\, p\, I \,a (1 - a)  \ \ \  {\rm with }\ \    I={e^2 \over h } V\ \ .
\end{equation}
%
Since the distribution Eq.~(2) gives rise to both a particle and a hole current, the prefactor in Eq.~(\ref{fano}) is 4 instead of the usual 2 (see inset of Fig.~\ref{fanopl}). 
%
\begin{figure}[t]
\includegraphics[width=0.8\columnwidth]{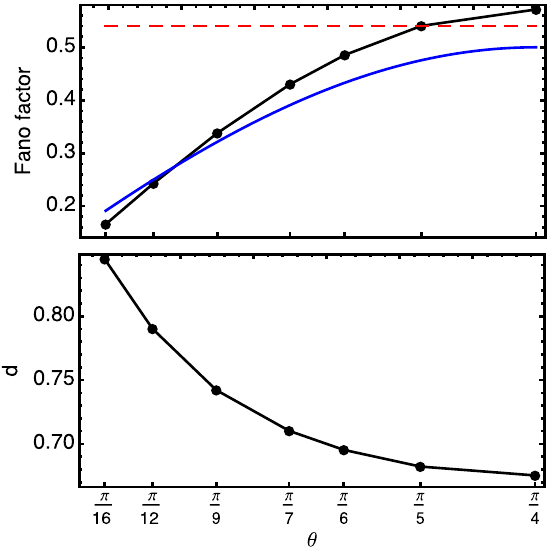}
\caption{(color online) Upper panel: Fano factor  as a function of the mixing angle for transmission $a=1/2$ of QPC1. (red dashed line) Fully equilibrated edge, $F$ is independent of interactions. (black dots) Full non-equilibrium situation. (blue line) Reference model of a diluted system of fractional charges ($F = (1/2) \sin 2\theta$). Lower panel: The dependence of the full non-equilibrium noise is calculated numerically and fitted by a function proportional to $(a(1-a))^d$ for different values of $\theta$. Black lines connecting the dots are a guide to the eye.} 
\label{varthetal}
\end{figure}
%
Defining a Fano factor $F=S/S_{\rm ref}$, we can make contact with the concept of fractional charges described in the introduction. Assuming  that for fractional charges the tunneling probability $p$ in Eq.~(\ref{fano}) is renormalized to $(e^*/e)p$ \cite{supplemental},  the Fano factor is given by $F=\sin2\theta/2$.   In  Fig.~\ref{varthetal} , the Fano factor is shown as a function of mixing angle for the specific transmission  $a=1/2$ of QPC1. For this value of $a$, there is a surprisingly good agreement between the value $e^*/e=(1/2) \sin 2 \theta$ and $F$ of the full non-equilibrium noise, suggesting that the Fano factor can indeed be interpreted as being due to formation of fractionalized charges  in the $\nu=2$ quantum Hall edge. 

We find that the zero frequency noise power depends in a singular way on $a$ in the limit  $a\ll1$, see also \cite{LeSu12a}.  To obtain the noise in this limit, the functional determinant can be approximated by its long time asymptotics (valid for $eV \tau \gg1$) 
$\bar{\Delta}_{\tau}(\delta) \simeq \exp(-|\tau|/(2\tau_{\phi}))$,
where the dephasing rate $\tau_{\phi}^{-1}= -(eV/2\pi) \log[1-4a(1-a)\sin^2(\pi \gamma_{\theta}/2)]$. Knowledge of 
$\bar{\Delta}_{\tau}(\delta)$ for large times allows to accurately calculate the distribution function of mode $(2u)$ for energies 
$\epsilon \ll e V$. However, for $a\ll1$ the distribution function only deviates from a step function on the scale $a e V$, 
such that the long time asymptotics allows an exact calculation of the distribution function. 
Using Eq.~(\ref{noise}) and taking the $a\ll1$  limit, we find $S(\omega \to 0) \simeq 8 p a \log(1/a)  \sin^2(\pi \gamma_{\theta}/2) eV (e^2/h\pi^2) \, $. This non-analyticity in $a$ explains the divergence in $S$ with $x_0$ found in \cite{Neder12} when calculating $S$ perturbatively in $a$.

A useful way to characterize the nonlinear dependence of experimentally measured  shot noise on the transmission probability $a$ of QPC1  is by fitting it to a function proportional to $(a(1-a))^d$ \cite{heiblum}. For the reference noise of Eq.~(\ref{fano}), d is trivially equal to unity. For "thermal"  noise with effective temperature $T^*$, one finds $d=0.5$.  For the full non-equilibrium noise, we find that its dependence  on  $a$ can be well fitted by the above power law, and that $d$  varies  from $d=0.85$ for $\theta =\pi/16$ to 
 $d=0.68$ for $\theta = \pi/4$, see Fig.~\ref{varthetal}. In this way, from knowledge of $d$ the mixing angle $\theta$ can be inferred, without using the Fano factor.

In summary,  due to the joint effect of interactions and non-equilibrium, the distribution function of an originally unbiased, zero temperature mode $(2u)$ interacting with a noisy mode $(1u)$
evolves towards a non-thermal steady state  that depends on the interaction strength in an characteristic way. 
Comparing the shot noise and Fano factor from our numerically exact 
calculation with a simple model of charge fractionalization, we find that the Fano factor can indeed 
be interpreted in terms of charge fractionalization in the $\nu=2$  quantum Hall edge. 
 
We would like to thank M.~Heiblum and H.~Inoue for valuable discussions, and acknowledge financial support by BMBF.  



\begin{widetext}

\section{Supplemental Material}  

In the first part of this supplemental material we present a derivation of shot-noise for the simple model of independent fractionalized charge pulses discussed in the main text. We proceed by considering an alternative derivation of the window function presented in the main text. The aim of this alternative derivation is to show that the form of the window function does not depend on the particular protocol used to switch on the interactions. We find that following a different protocol, the window function still describes two pulses of opposite signs propagating in the same direction. Hence all the results derived in the main text are exact and do not depend on using the quantum quench model instead of the non-equilibrium bosonization discussed in this supplemental material. 
\subsection{Shot-noise from the charge fractionalization model}
The physics of charge fractionalization in the $\nu=2$ quantum Hall (QH) state can be understood in the context of a simple charge fractionalization model for the setup of Fig.~(1) in the main text. The two chiral  channels on each edge of the sample co-propagate at different velocities $v_1$ and $v_2$. In the presence of a short range interaction $v_{12}$ between them, a pulse of charge $e$ injected into edge channel 1 at a first quantum point contact (QPC1) decomposes into a charge pulse and a neutral pulse as shown in Fig.~(\ref{cfract}). The charge and neutral mode pulses $(\tilde{\rho}_1,\tilde{\rho}_2)$ correspond to eigenmodes of the quadratic Hamiltonian (see main text),  and can be 
obtained from the transformation matrix 
\begin{figure*}[t]
\center{
\includegraphics[scale=1.1]{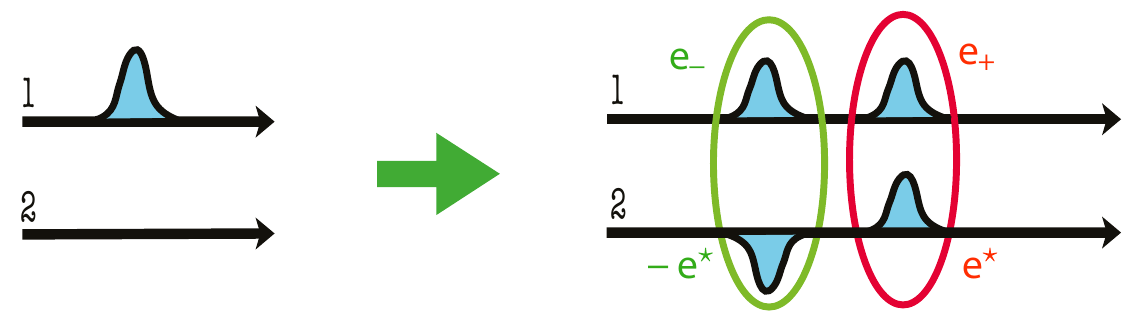}
}
\caption{(color online) Charge Fractionalization in a $\nu = 2$ QH state. A charge pulse initially injected in edge mode $1$ separates in a neutral (green) and charge (red) mode as a result of inter-channel interactions. The quasiparticles on edge mode $2$ have charges $e^*= \sin2\theta/2$, while the quasi particles on edge mode $1$ have charges $e_{\pm}=e/2 \pm \sqrt{e^2/4 -  (e^*)^2}$.}
\label{cfract}
\end{figure*}
\begin{equation} \label{supp:bog}  
M=\left(\begin{array}{cc}\cos\theta & \sin\theta \\\ -\sin\theta & \cos\theta\end{array}\right) \ \ \ {\rm as} \ \ \ 
\left( \begin{array}{c} \tilde{\rho}_1 \\ \tilde{\rho}_2 \end{array} \right) \ = \ M  \left( \begin{array}{c} {\rho}_1 \\ {\rho}_2 \end{array} \right)
\ \ \ .
\end{equation}
When injecting a unit pulse into edge channel 1, we hence find $\tilde{\rho}_1 = \cos \theta$ and $\tilde{\rho}_2 = \sin \theta$. 
After the two pulses have separated due to their different velocities, the fractionalized charges on the two edge channels  can be obtained from $M^{-1} (\tilde{\rho}_1,0)^T$ for the charge pulse and from $M^{-1} (0,\tilde{\rho}_2)^T$ for the neutral pulse.  In the former, a charge $e^* = (e/2) \sin 2\theta$ (where $\tan2\theta=v_{12}/(v_1-v_2)$ parametrizes the strength of interactions) travels in channel 2 and  $e_+= e/2 + \sqrt{e^2/4 -  (e^*)^2} $ in channel 1. In the neutral pulse, there is a charge $- e^*$ in channel 2 and a charge $e_- = e/2 - \sqrt{e^2/4 -  (e^*)^2} $ in channel 1, see Fig.~(\ref{cfract}). 

In  a next step, we would like to use the above argument for charge fractionalization  to derive an expression for the current noise in channel  two of the bottom edge due to partitioning at QPC2. 
Clearly, QPC2 can only transmit electrons. As there are pulses with a fractional charge $e^*$ impinging on it, the crucial step in this derivation is to assign the correct probability to a process where an impinging fractional pulse causes the tunneling of an electron through QPC2.  As shown in Fig.~(\ref{cfract}), edge mode $2u$  contains both $\pm e^*$ charges that we will consider as spatially well separated and hence uncorrelated with each other. As a consequence, we assume that both the distribution of impinging charges and the tunneling at QPC2 are  governed by Poisson statistics . Let us focus  on the $e^*$ charges arriving at QPC2 and denote the corresponding impinging current by $I_{\rm imp}$. Next,  we denote  the  current measured at contact $3$ by $I_3$; we emphasize  that this is a current of electrons since only electrons can tunnel at QPC2. 
The impinging and the measured currents are  related via the reflection probability $p$ at QPC2 as $I_{\rm 3}=p \, I_{\rm imp}$. The two currents can be generally expressed in terms of the  number of fractional pulses $N_{\Delta t,  e^*}$ for $I_{\rm imp}$ and the
number of electrons $N_{\Delta t, e}$ for $I_{3} $, which pass 
 through a reference point within a time interval $\Delta t$. Specifically,  
\begin{eqnarray} \label{eq:fracnoise}  
I_{\rm imp} &=& e^* \, \frac{ \langle N_{\Delta t,  e^*} \rangle}{\Delta t} \\ \nonumber
I_{3} &=& e  \, \, \frac{  \langle N_{\Delta t, e} \rangle}{\Delta t} \ \ . 
\end{eqnarray}
Note that here $\langle O(t) \rangle$ means a time average of the observable. Using the relation between the measured and the impinging current we find 
\begin{equation}\label{eq:fracnoise1}
 \langle N_{\Delta t, e} \rangle= \left( p \frac{e^*}{e} \right) \langle N_{\Delta t,  e^*} \rangle.
\end{equation}
This relation can be understood in terms of a renormalized  probability $p \to p \, e^*/e $ for a process in which an  impinging $e^*$ pulse 
causes the tunneling of an electron at QPC2. 

Next, we consider the auxiliary problem of computing the current noise at a QPC onto which a {\em currentless} noisy edge mode characterized by the double step distribution Eq.~(2) of the main text impinges.  As illustrated in the inset of Fig.~4 of the main text, such a  double step distribution corresponds to a particle current $I_p$ and a hole current $I_h$ of equal magnitude, both being equal to 
%
\begin{equation}
I_p \ =\ I_h \ =\  {e^2 \over h} V a(1 - a)  \ \ .
\end{equation}
%
Due to the assumption of Poisson statistics and in the limit of small backscattering $p$ at QPC2, each of them gives rise to a 
shot noise in $I_3$ of magnitude $2 e p I_{p/h}$. Adding up the contributions of particle and hole currents, and introducing the 
initial current $I = {e^2 \over h} V$, we find the reference noise for the auxiliary problem as 
%
\begin{equation}
S_{\rm ref} (\omega \to 0) \ = \ 4 e p I a (1-a) \ \ .
\end{equation}
%
In a last step, we recall that channel two of the upper edge carries fractionalized charges $\pm e^*$ and not electrons, such that the 
tunneling probability at QPC2 is renormalized according to $p \to p \, e^*/e $ as discussed above. Then,  the "fractionalization noise" in 
our model is given by $S = (e^*/e) S_{\rm ref}$, and hence the Fano factor by 
%
\begin{equation}    
F \equiv {S \over S_{\rm ref}} \ = \ {e^* \over e}  \ \ . 
\end{equation}

%
%
\subsection{Alternative derivation of the window function}
We consider the setup of Fig.(1) in the main text, where QPC1 drives edge mode 1 out of equilibrium. 
While we modeled the spatial dependence of the interaction in terms of a temporal dependence by using the quantum quench formalism  in the main text, here we faithfully take into account  the spatial structure of the interaction term by closely following  the formalism originally developed in ref. \cite{GuGeMi}. The two edge modes are described in the fermionic language by the following real time, chiral fermionic action 
\begin{equation}\label{action} 
S=\int_{x,t} \left\{ \psi^{\dagger}_{1u} \imath (\partial_t+v_{1} \, \partial_x) \psi_{1u} +\psi^{\dagger}_{2u} 
\imath (\partial_t+v_{2} \, \partial_x) \psi_{1u} - 2 \pi \, v_{12}(x)  \, \rho_{1u} \rho_{2u} \right\},
\end{equation}  
where  $v_1$ and $v_2$ denote the velocities of the two modes. The last term in Eq.~(\ref{action}) describes a density-density interaction of strength $v_{12}(x)$ between the two edge modes (having same chirality), which is switched on directly after QPC1. We note that in the present approach,  interactions are turned on as a function of space as depicted in the setup Fig.~(1). In contrast, in the main text the chirality of the model was used to mimic the spatial switching-on of the interactions by a temporal one as in \cite{KoCh}, and the purpose of this supplemental material is to explicitly demonstrate the equivalence of the two approaches. 
The density fields are defined as $\rho_i= \psi^{\dagger}_{i}\psi_{i}$. According to \cite{GuGeMi}, the non-interacting part of the chiral action (\ref{action}) can be bosonized  in Keldysh space as
%
\begin{figure}[t]
\includegraphics[scale=1.5 ]{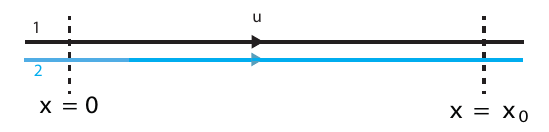}
\caption{(color online) Schematic of the top edge modes. In this schematic we depict the two edge modes between the two QPCs. QPC one is located at $x=0$ while QPC2 is located at $x=x_0$, where the equal space Green functions are evaluated. We consider switching on the interaction right after the first QPC. Due to the chirality of the model, this procedure does not change the calculations.}
\label{int}
\end{figure}
%
\begin{eqnarray}\label{bos0} \nonumber
S_{0,i}[\rho_i,\bar{\rho}_i]&=& \int_{x,t} \left\{ -\rho_i (\Pi_i^a)^{-1} \bar{\rho}_i- \imath \, \log Z_i[\bar{\chi}_i] \right\} \equiv S_{i,cl}+S_{i,q} \\ \label{0action} 
\imath \, \log Z_i[\bar{\chi}_i]&=& \sum_{n=2}^{\infty} \imath^{n+1} \bar{\chi}^n_i \, \mathscr{S}_{n,i}/ n! \\ \label{quantumS}
 \bar{\chi}_i&=&  (\Pi_i^a)^{-1} \bar{\rho}_i. \label{counting}
 \end{eqnarray}
After bosonization, the theory is expressed solely in terms of the density fields $\rho_i$, related to bosonic phase fields $\phi_i$ by $\rho_i= \partial_x \phi_i/ 2\pi$.  In Keldysh space, the fields $\rho_i$ and $\bar{\rho}_i$ are respectively the classical and quantum components of the charge density. In Eq. (\ref{0action}) we have separated the non interacting action in a sum of two terms: the classical one $S_{i,cl}$ (containing informations about the spectral properties) and the quantum one $S_{i,q} $.    
The quantum action is expressed in terms of a sum over the $n$ vacuum loops  $\mathscr{S}_{n,i}$  ($n=2$ corresponding to the RPA bubble) and $ \bar{\chi}_i$ is a quantum field that in the non-equilibrium bosonization formalism is interpreted as the counting field of full counting statistics \cite{GuGeMi}. The classical part of the action always involves a product of a quantum and a classical field. In the loop expansion the "coefficient" of the classical components are the advanced and retarded polarization functions (note that here we follow \cite{GuGeMi} and write only the advanced component, the two being related by complex conjugation). For the classical component of the effective action, RPA is exact as in conventional bosonization. Note that the vanishing of higher order loops corresponds to the absence of vertex corrections found by Dzyaloshinskii and Larkin \cite{DL}. 
The quantum part always contains a product of quantum fields and its coefficients,  $\mathscr{S}_{n,i}$, are the Keldysh part of the vacuum loops. For the quantum component, RPA is not exact and all loops must be considered. The advanced component of the polarization operator is found as (in energy momentum space)
\begin{equation}\label{polarization}
\Pi^a_i(q,\omega)=\frac{1}{2\pi}\frac{q}{v_i \,q-\omega+\imath\, 0^+}  \ \  .
\end{equation}
In the above expression, $i=1,2$ labels the two edge modes. 
Bosonization of the interacting part of the action can be achieved using the Hubbard-Stratonovich transformation in the particle-hole channel as explained in \cite{GuGeMi}. Integrating out the auxiliary field,  the interacting part of the action reads (all the fields are now evaluated for $x>0$)
\begin{equation}\label{intbos} 
S_{\rm int}[\rho,\bar{\rho}]= - \int_{x,t}  \pi \, v_{12} \left\{    \rho_1 \,  \bar{\rho}_2    +    \bar{\rho}_1 \, \rho_2    \right \}  \ \ .
\end{equation}
In this way we are left considering the effect of the inter-mode interaction, which is  purely classical and hence  does not affect the quantum action. Including everything, the bosonic action is given by 
\begin{equation}\label{actbos} 
S[\rho,\bar{\rho}]= S_{cl}[\rho,\bar{\rho}]+S_{q}[\bar{\rho}],
\end{equation} 
where the classical part $S_{cl}[\rho,\bar{\rho}]= S_{0,cl}+S_{\rm int}$.
\subsubsection{Green functions}
For the sake of being self-contained, we reproduce some steps of the derivation in \cite{GuGeMi} . 
We are interested in evaluating the Green function of mode $2u$. In the following, we suppress the index $u$ since we will always refer to the upper edge. The "lesser", equal space, Green function is defined as
\begin{equation}\label{greater1}
G^{<}_{2}(\tau)=\langle  \psi_{2,-}^{\dagger}(x_0,t_0+\tau) \psi_{2,+}(x_0,t_0) \rangle 
\end{equation}
Here we have explicitly mapped the fields on the Keldysh contour, and $\pm$ refers respectively to the upper/lower branch of the contour. The space coordinate $x_0$ refers to the point where the Green function is evaluated (the position of the QPC2 in our case). Using the bosonic representation of the fermionic fields, the Green function assumes the form 
\begin{equation}\label{greater2}
G^{<}_{2}(\tau)=\frac{1}{2\pi\alpha} \langle e^{-\imath \phi_{2,-}(x_0,t_0+\tau)}  e^{\imath \phi_{2,+}(x_0,t_0)} \rangle, 
\end{equation}
where $\alpha$ is a short distance cutoff on the scale of the magnetic length. At this point it is convenient to perform a Keldysh rotation to classical ($\phi$) and quantum ($\bar{\phi}$) components in order to make use of the bosonized action (\ref{actbos}) 
\begin{equation}\label{krotation}
\phi_{2,+}=\frac{1}{\sqrt{2}} (\phi_2+\bar{\phi}_2) \quad , \quad  \phi_{2,-}=\frac{1}{\sqrt{2}} (\phi_2-\bar{\phi}_2).  
\end{equation}  
After the Keldysh rotation, Eq~(\ref{greater2}) reads
\begin{eqnarray}\label{greater3}  
G^{<}_{2}(\tau)&=& \frac{1}{2\pi\alpha} \langle e^{-\frac{\imath}{\sqrt{2}} \left[\phi_{2}(x_0,t_0+\tau)-\bar{\phi}_2(x_0,t_0+\tau) \right]} \, \, \, e^{\frac{\imath}{\sqrt{2}} \left[\phi_{2}(x_0,\tau_0)+\bar{\phi}_{2}(x_0,\tau) \right] } \rangle \\ \nonumber 
&=& \frac{1}{2\pi\alpha} \int \mathscr{D}[\rho_1, \bar{\rho}_1] e^{\imath S_{0,\rm cl}[\rho_1,\bar{\rho}_1]} \,  e^{\imath S_{\rm q,1}[\bar{\rho}_1]} \, \\ \nonumber 
& &\times  \int \mathscr{D}[\rho_2, \bar{\rho}_2]
\, \, e^{\imath S_{0,\rm cl}[\rho_2,\bar{\rho}_2]} \, e^{\imath S_{\rm q,2} [\bar{\rho}_2]} \,  e^{\imath S_{\rm int}  [\rho,\bar{\rho}]} \,
\, e^{-\frac{\imath}{\sqrt{2}} \left\{ \phi_{2}(x_0,t_0+\tau)-\phi_{2}(x_0,t_0)-\bar{\phi}_{2}(x_0,t_0+\tau)-\bar{\phi}_{2}(x_0,t_0) \right\}   }.  
\end{eqnarray}  
In the second equality above, we have expressed the expectation value in terms of a path integral of bosonic density fields taken with the bosonized action of Eq.~(\ref{actbos}). We first consider the the classical components and introduce the source field
%
\begin{equation}
j_2(x,t) = \frac{1}{\sqrt{2}} \delta(x-x_0) \Big\{ \delta(t -t_0-\tau)- \delta(t-t_0) \Big\} \ \ 
\end{equation}
%
in order to evaluate the expectation value in Eq.~(\ref{greater3}) \cite{Wen}. Expressing the classical density fields in terms of their associated bosonic fields, the integrals over classical fields read
\begin{eqnarray}\label{classicint}  
& & \int \mathscr{D}[\phi_1, \phi_2] e^{-\imath \int_{x, t} \phi_1 \left\{-\frac{1}{2\pi} \partial_x [(\Pi_1^a)^{-1} \bar{\rho}_1] \right\}  }
e^{-\imath \int_{x, t} \phi_2 \left\{-\frac{1}{2\pi} \partial_x [(\Pi_2^a)^{-1} \bar{\rho}_2] + j_2 \right\}  }
e^{\imath S_{\rm int}  [\rho,\bar{\rho}]} \nonumber \\
& & =\delta \left(\frac{1}{2\pi} \partial_x [(\Pi_1^a)^{-1} \bar{\rho}_1  + \pi v_{12} \bar{\rho}_2] \right)
\delta \left(\frac{1}{2\pi} \partial_x [(\Pi_2^a)^{-1} \bar{\rho}_2   +  \pi v_{12} \bar{\rho}_1]    - j_2 \right).
\end{eqnarray}  
Using the expression 
\begin{eqnarray}\label{diag}
(\Pi_{i}^a)^{-1} &=& 2\pi \frac{v_i \, q-\omega+\imath 0^+}{q} \ \ ,
\end{eqnarray}
the Green function takes the form
\begin{eqnarray}\label{greater5}
G_2^<(\tau)&=&  \frac{1}{2\pi\alpha} \int \mathscr{D}[ \bar{\rho}_1] \,  e^{\imath {S}_{q,1} }    \int \mathscr{D}[\bar{\rho}_2] \, e^{ \imath S_{q,2} }  \, e^{\frac{\imath}{\sqrt{2}}   \left[ \bar{\phi}_{2}(x_0,t_0+\tau)+\bar{\phi}_{2}(x_0,t_0) \right ] }  \\ \nonumber
&\times&  
\delta \left[ (\partial_{t}+ v_1 \partial_x) \bar{\rho}_1 + {1 \over 2} v_{12} \partial_x \bar{\rho}_2          \right]
 \times \delta \left[ (\partial_{t}+ v_2 \partial_x) \bar{\rho}_2  + {1 \over 2} v_{12} \bar{\rho}_1   - j_2 \right].
\end{eqnarray}
To solve these coupled differential equations and to
make contact with the analysis presented in the main text, we perform a unitary transformation of the fields in order to bring the equilibrium part of the action into a diagonal form: $\bar{\eta}_{i}=\sum_{j} M_{ij} \bar{\rho}_j$, where the transformation matrix is chosen as
\begin{equation} \label{bog}  
M=\left(\begin{array}{cc}\cos\theta & \sin\theta \\\ -\sin\theta & \cos\theta\end{array}\right). 
\end{equation}
For this transformation, we assume that the interaction is switched on at the spatial point $x=0$, such that $v_{12}(x) = 
v_{12} \theta(x)$. In addition, we assume that the observation point $x_0$ is asymptotically far away from zero, $x_0 \gg {\hbar \over e V } v_{12}$. Then, we can ignore transients in the vicinity of $x = 0$, and solve the diagonal differential equations with the transformed source terms 
\begin{eqnarray}\label{sources}
\tilde{j}_1(x,t)&=& \frac{\sin\theta}{\sqrt{2}} \delta(x-x_0) \Big\{ \delta(t-t_0-\tau)- \delta(t  -t_0) \Big\} \\ \nonumber
\tilde{j}_2(x,t)&=& \frac{\cos\theta}{\sqrt{2}} \delta(x-x_0) \Big\{ \delta(t-t_0-\tau)- \delta(t  -t_0) \Big\}.
\end{eqnarray}
The equations of motion for the new density fields read    
\begin{equation}\label{motion1}
 (\partial_t+\tilde{v}_i \partial_x) \bar{\eta}_i(x,t)= j_i(x,t) \ \ , 
 \end{equation} 
where $\tilde{v}_i$ are the new velocities defined in the main text. Let us solve the equation of motion for $\bar{\eta}_1$ first.
We use the advanced Green function  
\begin{equation}\label{assgreen}
g_1^a(q,\omega)=\frac{-\imath}{\tilde{v}_1q-\omega+\imath 0^+}
\end{equation}
to obtain the solution for $\bar{\eta}_1$ 
\begin{eqnarray}\label{eta1}
\bar{\eta}_1(x,t) &=&\int_{q,\omega} \, g_1^a(q,\omega) \, j_1(q,\omega)= \frac{\sin\theta}{\sqrt{2}}  \int_{q,\omega} \frac{-\imath}{\tilde{v}_1q-\omega+\imath \, 0^+} \left\{ e^{\imath q(x-x_0)-\imath \omega (t-t_0-\tau) }- e^{\imath q(x-x_0)-\imath \omega (t-t_0) } \right\} \\ \nonumber
&=& \frac{-\sin\theta}{\sqrt{2}} \theta(x_0-x) \Big\{ \delta[(x-x_0)-\tilde{v}_1(t-t_0-\tau) ] - \delta[(x-x_0)-\tilde{v}_1(t-t_0) ] \Big\}.
\end{eqnarray}
Note that the analytical structure of the Green function is imposing a constraint on the real space dynamics, expressed by the step function appearing in the solution above. The solution for $\bar{\eta}_2$ is obtained in a similar way 
\begin{equation}\label{eta2}
\bar{\eta}_2(x,t) =\frac{-\cos\theta}{\sqrt{2}} \theta(x_0-x) \Big\{ \delta[(x-x_0)-\tilde{v}_2(t-t_0-\tau) ] - \delta[(x-x_0)-\tilde{v}_2(t-t_0) ] \Big\}
\end{equation}
At this point it is convenient to go back to the original basis, in which $S_q$ is known. The solution for the quantum density fields then read 
\begin{subequations}
\begin{eqnarray}
\label{tilderho1.eq}
\bar{\rho}_1(x,t) &=&  \frac{\sin2\theta}{2\sqrt{2}} \theta(x_0-x) \Big\{ \delta[(x-x_0)-\tilde{v}_1(t-t_0) ]-\delta[(x-x_0)-\tilde{v}_1(t-t_0-\tau) ] \\ \label{tilderho2}
&+& \delta[(x-x_0)-\tilde{v}_2(t-t_0-\tau) ] - \delta[(x-x_0)-\tilde{v}_2(t-t_0) ] \Big\} \\ \nonumber
\bar{\rho}_2(x,t) &=&  \frac{-\sin^2\theta}{\sqrt{2}} \theta(x_0-x) \Big\{ \delta[(x-x_0)-\tilde{v}_1(t-t_0-\tau) ]-\delta[(x-x_0)-\tilde{v}_1(t-t_0) ] \Big\} \\ \nonumber 
&-& \frac{\cos^2\theta}{\sqrt{2}} \theta(x_0-x) \Big\{\delta[(x-x_0)-\tilde{v}_2(t-t_0-\tau) ] - \delta[(x-x_0)-\tilde{v}_2(t-t_0) ] \Big\}.
\end{eqnarray}
\end{subequations}
Since $\bar{\rho}_2$ is at equilibrium, the integrals over the quantum fields yield an equilibrium Green function $G_0^<(\tau)$ multiplied
by a normalized functional determinant, which only depends on the scattering phase for the non-equilibrium mode $\bar{\rho}_1$, such that 
%
\begin{eqnarray}
G_2^< (\tau) & = & G_0^<(\tau) \, \bar{\Delta}_\tau[\delta] \ \ , 
\end{eqnarray}
%
with 
%
\begin{equation}
G_0^<(\tau) = \frac{1}{2\pi} \frac{1}{ (-\imath \,\tilde{v}_1 \,\tau + \alpha)^{\sin^2\theta} } \frac{1}{(-\imath \, \tilde{v}_2 \, \tau + \alpha)^{\cos^2\theta}}
\end{equation}
%
and 
\begin{equation}\label{fredh}
\bar{\Delta}_{\tau}[\delta]=\frac{\det[1+(e^{-\imath \delta_{\tau}}-1)f(\epsilon)] }{\det[1+(e^{-\imath \delta_{\tau}}-1)f_0(\epsilon)]},
\end{equation} 
where $f(\epsilon)$ is the double step Fermi distribution function given in Eq.~(2) of the main text and $f_0(\epsilon)$ is the zero temperature equilibrium distribution that we use as a normalization. Here, we are interested in determining the scattering phase $\delta_{\tau}$  defined as \cite{GuGeMi}  
\begin{equation}\label{phdef}
\delta_{\tau}(t_0) =  \sqrt{2} \int_{-\infty}^\infty \! \! d \tilde{t}\,  \, \overline{\chi}_1 (v \tilde{t}, \tilde{t} - t) \ \ . 
\end{equation}
By comparing our quantum density Eq.~(\ref{tilderho1.eq}) to that in Eq.~(59) in Ref.~\cite{GuGeMi}, from the window function in 
Eq.~(64) in Ref.~\cite{GuGeMi} we find the window function relevant for our problem
\begin{eqnarray} \label{intph}
\delta_{\tau}(t)&=& 2 \pi  \, {e^* \over e} \, \omega_\tau(t, t_0 - x_0/\tilde{v}_{1,2})
\end{eqnarray}
with 
\begin{eqnarray}
\omega_\tau(t, t_0 - x_0/\tilde{v}_{1,2}) & = &   \theta[ t_0 - x_0/\tilde{v}_1 -t] - \theta[t_0 - x_0/\tilde{v}_1 - (t + \tau)] + 
\theta[t_0 - x_0/\tilde{v}_2 - (t + \tau)]-\theta[t_0 - x_0/\tilde{v}_2 - t]   .
\label{windowfunction.eq}
\end{eqnarray}
We note that in the framework of \cite{GuGeMi} the functional determinant is evaluated with respect to time $t$, while in the main text it is evaluated with respect to time $t_0$. Given that the window function in Eq.~(\ref{windowfunction.eq}) only depends on the combination $t_0 -t$, it is clear that this difference has no qualitative consequences. Neglecting the parameters $x_0$, $t_0$ in the above equation and focusing on the dependence on the two times $t$, $\tau$, we find agreement with the dependence on times $t_0$, $\tau$ of the window function in the main text, with the only difference being an overall minus sign. Due to the fact that the determinant with a two-pulse window function separates into a product of determinants with one-pulse window functions, this overall minus sign is immaterial, and we find full agreement between the Green function computed using the quantum quench approach and the Green function computed using the functional approach of \cite{GuGeMi}.   Hence all the results derived in the main text are exact and do not depend on using the quantum quench model instead of the non-equilibrium bosonization discussed in this supplemental material.

\end{widetext}

\end{document}